# Near- and mid-infrared excitation of ultrafast demagnetization in a cobalt multilayer system

Katherine Légaré[1], Guillaume Barrette[1], Laurent Giroux[1], Jean-Michel Parent[1], Elissa Haddad[1], Heide Ibrahim[1], Philippe Lassonde[1], Emmanuelle Jal[2], Boris Vodungbo[2], Jan Lüning[3], Fabio Boschini[1], Nicolas Jaouen[4], François Légaré[1,*]

[1]Advanced Laser Light Source (ALLS) at Institut National de la Recherche Scientifique, Centre Énergie Matériaux Télécommunications, 1650 Boulevard Lionel-Boulet, Varennes, Québec J3X 1P7, Canada

[2]Sorbonne Université, CNRS, Laboratoire de Chimie Physique–Matière et Rayonnement, LCPMR, 75005 Paris, France

[3]Helmholtz-Zentrum Berlin für Materialien und Energie, 14109 Berlin, Germany

[4]Synchrotron SOLEIL, L'Orme des Merisiers, 91192, Gif-sur-Yvette, France

*francois.legare@inrs.ca

**Abstract:** In the last few decades, ultrafast demagnetization elicited by ultrashort laser pulses has been the subject of a large body of work that aim to better understand and control this phenomenon. Although specific magnetic materials' properties play a key role in defining ultrafast demagnetization dynamics, features of the driving laser pulse such as its duration and photon energy might also contribute. Here, we report ultrafast demagnetization of a cobalt/platinum multilayer in a broad spectral range spanning from the near-infrared (near-IR) to the mid-infrared (mid-IR), with wavelengths between 0.8 μm and 8.7 μm. The ultrafast dynamics of the macroscopic magnetization is tracked via the time-resolved magneto-optical Kerr effect. We show that the ultrafast demagnetization of the sample can be efficiently induced over that entire excitation spectrum with minimal dependence on the excitation wavelength. Instead, we confirm that the temporal profile of the pump excitation pulse is an important factor influencing ultrafast demagnetization dynamics.

## 1   Introduction

Ultrafast demagnetization is the phenomenon by which a sample's magnetic order is quenched on a sub-picosecond timescale upon excitation with an ultrafast light pulse. The first experimental evidence of ultrafast demagnetization dates back to 1996, and it immediately sparked both fundamental and technological interests [1]. Indeed, such an ultrafast, sizable change in magnetic order is promising for future magnetic data storage devices which require fast control over magnetization in order to improve read/write times [2]. As a result, ultrafast magnetism has grown into a vibrant research field. Recent discoveries such as all-optical switching [3,4], the manipulation of topological spin textures [5,6], and ultrafast spintronics [7] are bringing us ever closer to integrating ultrafast magnetization dynamics in future technologies.

Beyond these technological advancements, research is still ongoing to uncover the microscopic mechanisms responsible for ultrafast demagnetization [8]. In the last three decades, several mechanisms have been suggested to account for the loss of angular momentum during the demagnetization process. They can be divided in two categories: spin transport through the sample, which leaves a deficit of angular momentum in the excited region [9–15], and momentum transfer from the spin system to other degrees of freedom via different processes, such as electron-magnon scattering [16–18], electron-photon interactions [19], and electron-phonon spin-flip scattering [20,21].

In an effort to unveil what mechanisms underlie the ultrafast demagnetization, various samples with disparate electronic and magnetic properties have been investigated under different experimental conditions [22]. Independently from the mechanism responsible for the ultrafast quench of the magnetic order, it is well established that the first and pivotal step for the demagnetization process in metallic systems is the quasi-instantaneous transfer of energy from the pump pulse to the electronic bath [23–25]. An open question is whether and how different light-induced redistribution of the charges in the energy-momentum phase space impact the demagnetization dynamics. This can be investigated, for instance, by pumping the system using different excitation wavelengths. Nevertheless, most studies have exclusively used readily available near-infrared (near-IR) pump pulses to elicit ultrafast demagnetization, with only a few investigations in the extreme ultraviolet [26,27] and terahertz spectral ranges [28–32]. To our knowledge, the only attempt to pump ultrafast demagnetization in the mid-infrared (mid-IR) has been reported in a conference paper by Zagdoud *et al.* [33].

In this work, we explore how the wavelength of the pump pulse can affect ultrafast demagnetization dynamics in a ferromagnetic sample. Previous studies on insulated thin films, where spin transport is minimized, show no wavelength-dependent demagnetization dynamics [31,34]. Conversely, the effect of the pump wavelength is apparent in bulk materials and heterogeneous structures [35–38]. In particular, Cardin *et al.* showed that the demagnetization becomes more efficient as the pump wavelength increases from 0.4 µm to 1.8 µm in a cobalt/platinum multilayer sample [38]. In this specific case, the wavelength dependence may have been amplified by the distribution of the absorbed energy within the sample structure; indeed, the portion of the pump energy that is directly absorbed by the magnetic multilayers rather than by the capping aluminium layer is also wavelength dependent. Here, we extend the pump wavelength towards the mid-IR spectral range, where the fraction of energy absorbed by the magnetic system is mostly constant. There, we also avoid promoting inter-band transitions, which depend intrinsically on the material. This experimental strategy allows us to reach more general conclusions about magnetic systems. By probing ultrafast magnetic dynamics of a Co/Pt multilayer system with out-of-plane magnetization through the polar magneto-optical Kerr effect (P-MOKE), we show that ultrafast demagnetization can be efficiently induced throughout the entire range of studied pump wavelengths (from 0.8 µm to 8.7 µm). While demagnetization dynamics are weakly wavelength dependent within the near-IR, an effect that has previously been reported for near-IR to ultraviolet regions [35,36,38], we do not find a significant wavelength dependency in the mid-IR range and the temporal shape of the pulse plays indeed a significant role.

## 2  Method

The experiments are performed on a Co/Pt multilayer sample caped by an Al layer and deposited on an Si substrate. The Si/Ta$_{3nm}$/Pt$_{2nm}$/[Co$_{0.6nm}$/Pt$_{0.8nm}$]$_{x20}$/Al$_{3nm}$ multilayer is grown by DC magnetron sputtering. The magnetization of the sample is probed using P-MOKE and a pump-probe scheme allows to gather time-resolved information. The setup, shown in Fig. 1, is similar to the P-MOKE setup presented in Légaré *et al.* [39]. Both the pump and probe pulses originate from a single Titanium-Sapphire laser system available at the Advanced Laser Light Source user facility delivering 800 nm, 45 fs pulses at a repetition rate of 100 Hz. In the laser system, the beam is separated in two parts that are independently amplified through chirped-pulse amplifiers, leading to a high energy (20 mJ pulse energy) and a low energy (5 mJ pulse energy) output. As explained below, the higher energy beam is used to generate wavelengths up to 8.7 um. The lower energy beam is further divided in two parts which later become the pump and the probe.

In the probe line, the beam contains less than 200 µJ of energy per pulse and it is frequency doubled through a 100 µm thick, type I β-barium borate (BBO) crystal inserted in the beam. The use of 400 nm light reduces the state-filling effects that can affect ultrafast MOKE measurements [40,41], and it ensures that the pump penetrates deeper than the probe in the metallic sample so that the pumped volume is larger than the probed volume for all wavelengths. Two dichroic mirrors and a band-pass filter are used to remove the fundamental wavelength from the beam. A glass plate placed at the Brewster angle sets the polarization state of the probe beam to the s-polarization to facilitate the retrieval of the Kerr rotation [39].

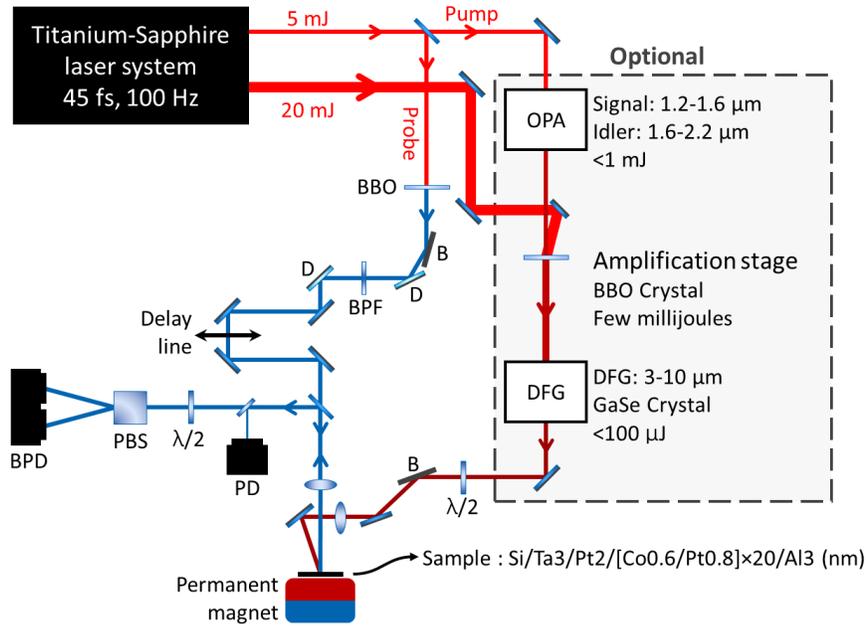

**Figure 1 – Time-resolved MOKE experimental setup. The components are: (BBO) β-barium borate crystal used for frequency doubling, (B) Glass or silica plate placed at the Brewster angle, (D) dichroic mirror, (BPF) band-pass filter, (λ/2) half-waveplate, (PBS) polarization beamsplitter, (BPD) balanced photodetectors, (PD) Photodetector, (OPA) optical parametric amplifier, (DFG) difference frequency generation stage.**

At the sample surface, the 400 nm probe beam is focused down to a diameter of 50 μm and each pulse has an energy <<1 μJ. A permanent magnet placed behind the sample ensures that the sample's magnetization is saturated by applying a field of ~300 mT. The reflected beam is then sent to a half-waveplate, a polarization beamsplitter, and balanced photodetectors. Every measurement is repeated with the sample's magnetization saturated in the opposite direction ($I^+(t)$ and $I^-(t)$) and the difference between those measurements is taken to be proportional to the magnetization: $\frac{\theta(t)}{\theta_0} = \frac{I^+(t)-I^-(t)}{I^+(t<0)-I^-(t<0)} = \frac{M(t)}{M_0}$. This scheme allows to measure the rotation of the polarization due to MOKE with a high signal-to-noise ratio [39]. An additional detector is used to monitor the sample's reflectivity and ensure that the ultrafast rotation of the polarization is solely due to magnetic dynamics [42]. This monitoring also allows to compensate for fluctuations of the probe pulse energy.

In the pump line, the wavelength must be tuned over a wide spectral range in order to excite the sample with near-IR to mid-IR pulses. To achieve this, several non-linear stages have been implemented. A commercial optical parametric amplifier (HE-TOPAS, Light Conversion, Inc.) is used to generate wavelengths between 1.2 µm and 2.1 µm. To reach the mid-IR, the OPA output is first mixed with the high-energy output beam of the laser system in a BBO crystal for further amplification (see Thiré *et al.* [43]). Then, a GaSe crystal is used to mix the OPA's signal and idler in a difference frequency generation (DFG) scheme to extend the wavelength up to 8.7 µm.

A half wave-plate followed by a silica plate at Brewster angle are inserted into the beam to control the pump pulse energy. The beam is then focused on the sample at near normal incidence (less than 10°) by a parabolic mirror. The sweeping effect adds a maximum of 30 fs smear to the temporal resolution of the measured dynamics. The focusing mirror is chosen such that the pump beam size on the sample is at least 3.5 times larger than the probe beam.

For each wavelength, time-resolved MOKE measurements are performed at several pump fluences. The measured quantity is the relative rotation of the polarization with respect to the saturated sample, $\theta(t)/\theta_0$. This quantity is commonly taken to be proportional to the relative magnetization amplitude $M(t)/M_0$ [44], although there is some controversy about the legitimacy of this relation for rapidly evolving systems [40,45,46]. The ultrafast demagnetization dynamics are captured by a bi-exponential fit function [47]:

$$\frac{M(t)}{M_0} = G(t) \otimes \left[ 1 - H(t) \left[ B \left( 1 - e^{\left(-\frac{t-t_0}{\tau_1}\right)} \right) e^{\left(-\frac{t-t_0}{\tau_2}\right)} + C \left( 1 - e^{\left(-\frac{t-t_0}{\tau_2}\right)} \right) \right] \right], \quad (1)$$

where $G(t)$ is a Gaussian function which represents the temporal resolution of the experiment and the Heaviside function $H(t)$ describes the abrupt drop of magnetization upon excitation of the sample by the pump. The pump and probe pulses are synchronized on the sample at the delay $t = t_0$ and $\tau_1$ and $\tau_2$ describe the durations of the magnetization drop and recovery, respectively. As described in [38], the parameters $B$ and $C$ roughly represent the maximum magnetization quenching and the remaining quenching after the partial recovery of magnetization, i.e. several picoseconds after excitation (Fig. 2). In the analysis, mainly $B$ and $C$ are used to compare magnetization dynamics obtained in different experimental conditions [38]. It should be noted that the slow recovery of magnetization, which is driven by lattice thermalization and occurs on the picosecond-to-nanosecond timescale, is not considered here as it falls outside the scope of this work.

Several effects contribute to the width of $G(t)$, such as the pump and probe pulse durations, the geometry of the beam paths, and the intrinsic duration associated to the onset of ultrafast demagnetization, which may not be instantaneous. The width of $G(t)$ is left as a free parameter for the numerical optimization of the fit function.

## 3  Results and discussion

Multiple demagnetization curves were measured with varying pump fluence for eight different pump wavelengths between 0.8 µm and 8.7 µm. An excerpt of the results is shown in Fig. 2. From these curves, some notable conclusions can already be made. Clearly, ultrafast demagnetization is achievable using mid-IR pump pulses, and the dynamics are not dramatically different from those commonly measured upon near-IR excitation for this sample. In all cases, magnetization drops on the femtosecond timescale and partially recovers on the picosecond timescale, reaching a plateau that depends on the pump fluence.

Typically, an absorbed fluence of a few mJ/cm$^2$ is required to efficiently induce demagnetization in Co/Pt multilayers [38,48]. In the experiment presented here, the exact pump fluences are difficult to assess because of the very low pump pulse energies and due to the challenges associated with the precise characterisation of the pump size on the sample, especially in the mid-IR. The *incident* pump fluence is estimated to tens of mJ/cm$^2$ for all pump wavelengths with a maximum of ~100 mJ/cm$^2$ at 8.7 µm. Since short wavelengths are more readily absorbed by the Co/Pt sample, a lower incident fluence is necessary to trigger ultrafast demagnetization from a near-IR pump. Indeed, using a multilayer modelling software [49], we estimate that 25% of the pump energy is absorbed at 0.8 µm versus 6% at 8.7 µm. Thus, the *absorbed* fluences used in this experiment are of the order of a few mJ/cm$^2$ for all pump wavelengths.

Given the challenges associated with the accurate estimate of the absorbed pump fluence, our experimental strategy consists in using either the parameter $\tau_2$ or the parameter $C$ from equation (1) as a gauge of the absorbed pump energy, as described in Cardin *et al*. [38]. Typically, the fast recovery time $\tau_2$ increases linearly with the pump fluence, however it is also affected by the temporal profile of the pump pulse [50,51], which is a concern since the pump pulse duration is wavelength-dependant in this work (see section 3.2). The parameter $C$, which describes the remaining quenching after partial magnetization recovery at long time delays, is therefore preferred. Indeed, several picoseconds after the pump pulse has left the sample, the remnant demagnetization reflects the overall energy deposited into the sample [50]. In the following, the ultrafast demagnetization dynamics are examined as a function of $C$ for each pump wavelength.

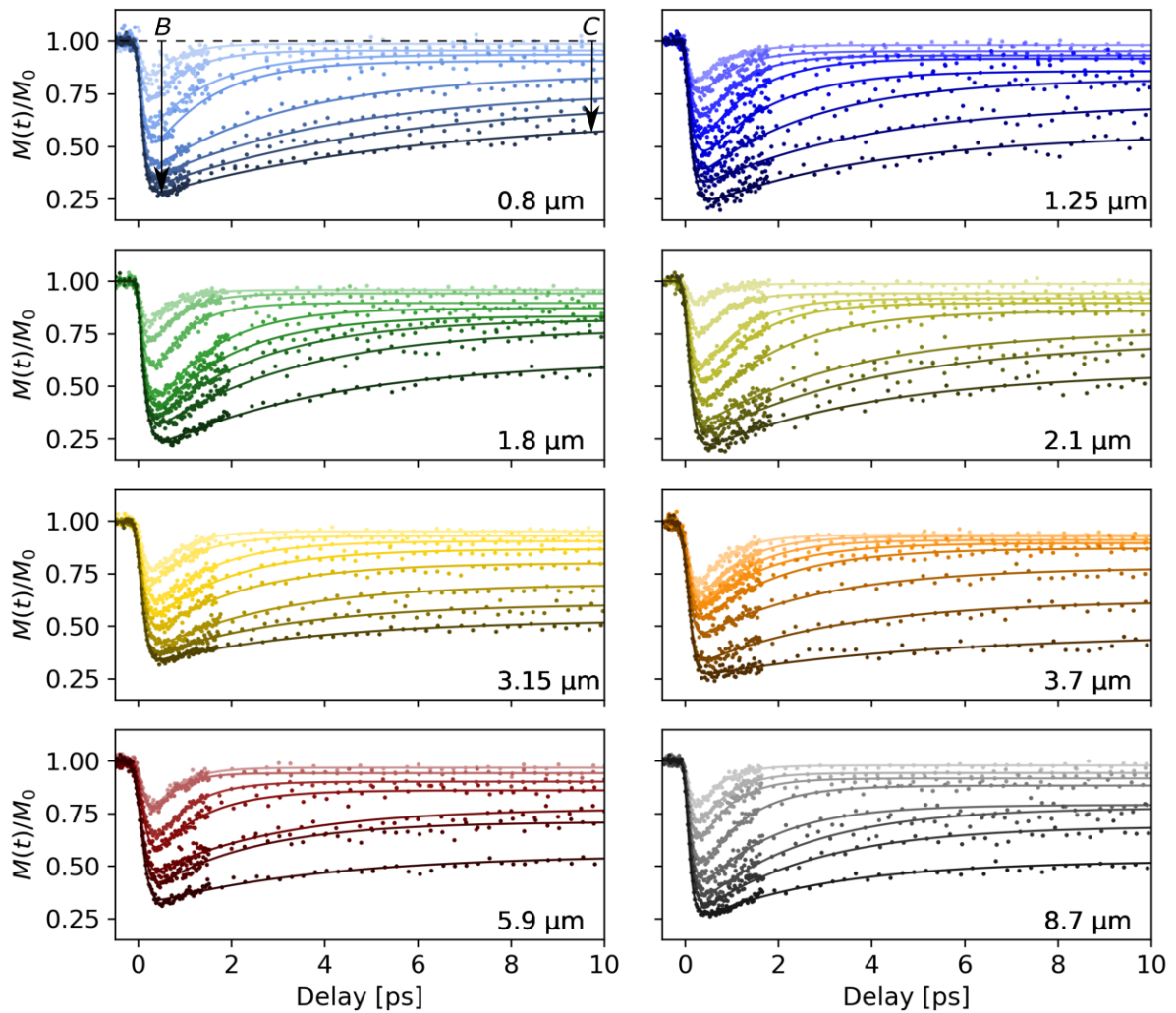

Figure 2 – Examples of ultrafast demagnetization curves measured with different pump fluences and for pump wavelengths of 0.8 µm, 1.25 µm, 1.8 µm, 2.1 µm, 3.15 µm, 3.7 µm, 5.9 µm and 8.7 µm, as indicated. The experimental data is fitted by equation (1). For each wavelength, the transition from light to dark curves corresponds to an increase of the pump fluence. As expected, larger pump fluences result in a stronger magnetization quenching at short delays ($B$) and long delays ($C$).

## 3.1 Deagnetization time

The characteristic demagnetization time describes how long it takes for the magnetization to drop towards its lowest point. It can be represented by the parameter $\tau_1$ of equation (1). In the literature, however, this quantity is most often defined as the time required for the magnetization quenching to reach $(1 - e^{-1})$ of its maximum [48]. To facilitate comparison with previously studied systems, this definition is adopted here.

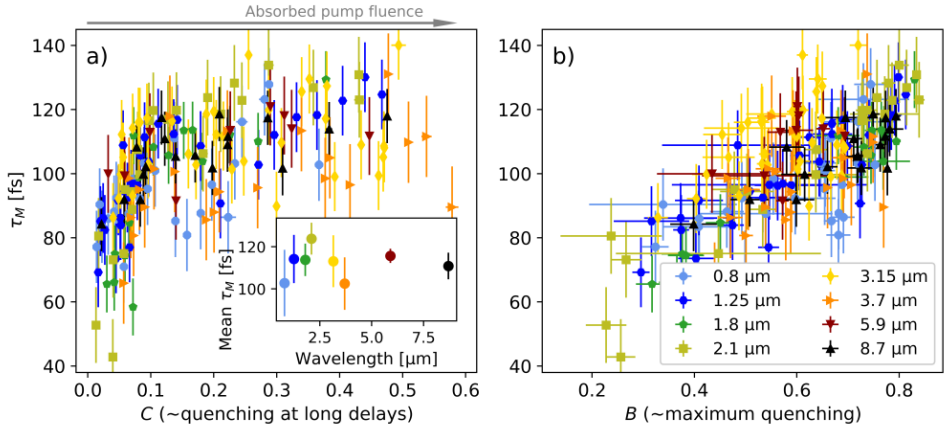

**Figure 3** – (a) Characteristic demagnetization time $\tau_M$ as a function of $C$ for all considered pump wavelengths and fluences. The color and shape of the data points correspond to the different pump wavelengths as per the legend shown in (b). Error bars correspond to a variance of one standard deviation on the numerical parameters estimate. Inset is the mean value of $\tau_M$ when $C > 0.15$ for each pump wavelength with error bars corresponding to one standard deviation. (b) Characteristic demagnetization time as a function of $B$, which roughly corresponds to the maximum magnetization quenching.

Fig. 3(a) shows the evolution of the characteristic demagnetization time $\tau_M$ as a function of parameter $C$. For each pump wavelength, the wide distribution of the data points is explained in parts by experimental artifacts such as fluctuations of the pump pulse energy and in parts by the uncertainty of the numerically fitted parameters. Globally, however, $\tau_M$ increases quickly at low $C$ (i.e. low pump fluence) before stabilizing when $C$ becomes large. Interestingly, the data clouds of each wavelength overlay each other. This means that the pump wavelength does not have a significant effect on the demagnetization time. Moreover, the inset of Fig. 3(a) shows that for large values of $C$, the averaged characteristic demagnetization time stays close to 110 fs for all pump wavelengths. This result, which had already been demonstrated for near-IR pumps [38], endures also in the mid-IR spectral range.

Plotting $\tau_M$ as a function of $B$ (Fig. 3(b)) reveals a slowly ascending slope. This is predicted by the spin-flip electron-electron scattering demagnetization model introduced by

Koopmans *et al.* [21], which also agrees with a small value of $\tau_M$ (~100 fs) for a strongly excited Co/Pt sample due to the enhanced spin-orbit coupling brought by the Pt layers [52]. It should be noted that other models, such as electron-magnon scattering, are also compatible with this data [17].

**3.2 Maximum magnetization quenching**

A few studies have reported data that shows an enhancement of the ultrafast magnetization quenching as a function of the pump wavelength in the near-IR [35,36,38]. Here, we look at how the maximum magnetization quenching ($B$ parameter) varies with the wavelength for a given value of the absorbed pump fluence up to the mid-IR. In Fig. 4(a), the progression of the maximum magnetization quenching as a function of the absorbed pump fluence is represented through the parameters $B$ and $C$ for a few selected pump wavelengths. Corresponding examples of the demagnetization curves measured with a pump fluence set so that $C = 0.3$ are shown in Fig. 4(b). Evidently, the demagnetization dynamics evolve with the different experimental conditions, but they do not seem to change monotonically with the pump wavelength.

To push the analysis further, a linear regression is applied to the curves shown in Fig. 4(a) for $C > 0.15$. Then, an extrapolated value of $B$ for $C = 0.3$ is calculated and plotted against the pump wavelength in Fig. 4(c). Although we report an enhancement of the maximum magnetization quenching when the pump wavelength increases from 0.8 to 1.8 µm (near-IR range), it is reduced in the 3.15 to 5.9 µm range. Surprisingly, the maximum magnetization quenching at 8.7 µm is comparable to that observed in the near-IR range.

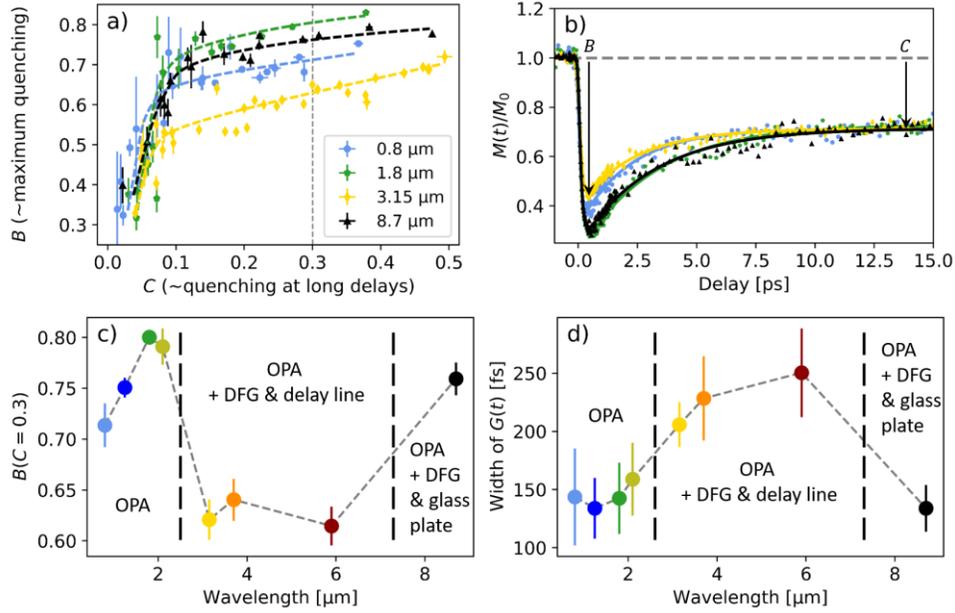

**Figure 4 –** (a) Parameter $B$ as a function of parameter $C$ for some of the studied pump wavelengths. The dashed lines only serve as a guide to the eye. (b) Demagnetization curves obtained when the pump fluence is set as to reach $C = 0.3$ for the pump wavelengths presented in (a). (c) Calculated maximum quenching $B$ when $C = 0.3$ for all pump wavelengths. The vertical dashed lines show the separation between the 3 different regimes. (d) Averaged full-width at half-maximum of $G(t)$ obtained from the numerical fitting of every demagnetization curve for each pump wavelength.

In an effort to understand this non-monotonic behaviour of the maximum magnetization quenching with respect to $\lambda$, we note that the three regimes identified in Fig. 4(c) and (d) can be correlated with different experimental approaches for the generation of the pump pulses. Rather than being the direct result of the pump wavelength, this could indicate that the variations of $B$ are instead linked to the temporal profile of the pump pulses.

As shown in Fig. 1, in the near-IR, the pump is generated from a commercial OPA leading to pulse durations of ~50 fs (characterized with second-harmonic-generation frequency-resolved optical gating). For the intermediate wavelengths, an additional amplification step, a DFG crystal and a delay line used to synchronize the signal and idler from the OPA in the crystal are added. At 8.7 µm, the delay line is replaced with a single dispersive glass plate. Considering the sensitive nature of the phase-matching that must be optimized for each non-linear process, these different methods can lead to different temporal characteristics of the pulses for each pump wavelength. An imperfect phase-matching could result in longer pulses, the transfer of energy towards a pedestal, or the creation of post-pulses. All of these effects can influence the ultrafast magnetization dynamics [50,53].

Despite the challenges associated to the characterization of the mid-IR pump pulses, which are highly sensitive to environmental conditions, their duration was estimated to 140±60 fs using frequency-resolved optical switching (FROSt) [54]. Additionally, it is possible to extract some information about the pump pulses from the numerical fit of the ultrafast demagnetization curves. Fig. 4(d) shows the full-width half maximum of $G(t)$ for each pump wavelength. The same three regimes as in Fig. 4(c) are visible; $G(t)$ is larger in the intermediate regime, which indicates that the pump pulses are either significantly longer or that they contain pedestals and post-pulses which the model is unable to properly extract.

It has been shown that ultrafast demagnetization dynamics driven by a long pump pulse can be empirically fitted by the convolution between the pump temporal profile and dynamics obtained from a short pump pulse [50,51]. The consequence of stretching the pump pulse duration is to lower the measured maximum quenching and lengthen the characteristic demagnetization time. In this work, although the former effect is observed for the intermediate regime (see Fig. 4(c)), the characteristic demagnetization time remains constant over all experimental configurations. Therefore, it is most likely that the pump pulses exhibit complex temporal profiles that contain post pulses rather than merely being stretched into longer Gaussian pulses. As demonstrated in Bühlmann *et al.* [53], demagnetization from two consecutive pump pulses can indeed affect picosecond dynamics without influencing the characteristic demagnetization time. In that case, it should be noted that equation (1) is not a valid representation of the demagnetization dynamics since the unique characteristics of the pump pulses are not taken into account.

As shown in Fig. 4(d), for pump wavelengths of 0.8, 1.25, 1.8, 2.1 and 8.7 μm, the width of $G(t)$ is mostly constant. It can therefore be assumed that the temporal profiles of the pump pulses are nearly identical for these conditions. As a result, one could expect the near-IR trend of increasing $B$ for a fixed $C$ to continue at 8.7 μm, but this it is not the case. In Fig. 4(c), the maximum magnetization quenching reached with a pump of 2.1 μm is comparable to the one at 8.7 μm. Therefore, it becomes clear that the effect of the wavelength on the demagnetization dynamics is both weak and non-monotonic.

## 4  Conclusion

To conclude, efficient ultrafast demagnetization is demonstrated over a broad range of excitation wavelengths from the near-IR to the mid-IR. Overall, we show that the ultrafast demagnetization dynamics of the Co/Pt multilayer sample depend only weakly on the pump pulse wavelength and that the effect is non-monotonic. Indeed, while longer wavelengths lead to slightly more efficient demagnetization within the near-IR spectral region, a trend that as also been shown to continue the ultraviolet region [35,36,38], this progression is not maintained up to a pump wavelength of 8.7 µm. Additionally, the temporal profile of the excitation laser pulses is identified as an important parameter that may influence the demagnetization dynamics. Further investigations on the role of the rate of excitation on ultrafast demagnetization are required to better understand this result. Since the spectrum investigated here is so large (from photons of 0.14 eV to 1.55 eV), it is also possible that the data covers a change of excitation regime from intra- to interband transitions. A shift towards much longer pump wavelength coupled with a better control of the pulse's temporal shape would allow to verify whether such a regime change can contribute the non-monotonic wavelength scaling of demagnetization dynamics with the wavelength that we observe here.

Finally, in conjunction with results published in the last few years, a pattern is starting to emerge; the excitation wavelength only seems to affect the demagnetization dynamics when non-local effects are present in the sample, which is the case for bulk or multilayer samples [35–38]. When non-local effects can be neglected, ultrafast magnetic dynamics are independent of the transient electron distribution [31,34]. As research turns to more complex systems that are promising for future technological applications of ultrafast magnetics, such development may inspire novel approaches for characterising and modelling magnetization dynamics and contribute to a better understanding of this ultrafast phenomenon.

## Acknowledgments

The authors acknowledge financial support from Ministère de l'Économie, de l'Innovation et de l'Énergie – Québec, PROMPT – Québec, the Canada Foundation for Innovation, the Fonds de recherche du Québec – Nature et technologies (FRQNT), and the Natural Sciences and Engineering Research Council of Canada (NSERC). K.L., J.-M. P. and E.H. acknowledge financial support from NSERC. The authors are grateful for the financial support received from the MEDYNA ANR-20-CE42-0012, as well as for the contribution of Renaud Delaunay (Sorbonne Université) in growing the sample.